\documentclass[twocolumn]{webofc}
\usepackage[varg]{txfonts}   % Web of Conferences font
\usepackage{hyperref}
\usepackage{amsmath}
\usepackage[font=small,labelfont=bf,justification=justified,format=plain]{caption}
\usepackage{float}
\usepackage{url}
\hypersetup{colorlinks=true,citecolor=blue,urlcolor=blue,linkcolor=blue}
\newcommand{\Rcore}{R_{\mathrm{core}}}
\newcommand{\Rstar}{R}                      % stellar radius
\begin{document}

\title{Exploring the internal structure of a neutron star and the associated magnetic fields aided by the mass-radius relationship
%Exploring neutron star crusts, cores, and magnetic fields using the mass-radius relation
}

\author{\firstname{Abriana} \lastname{Lyda}\inst{1}\fnsep\thanks{\email{arlyda@mit.edu}} \and
        \firstname{Prajwal} \lastname{MohanMurthy}\inst{1}\fnsep\thanks{\email{prajwal@alum.mit.edu}}}
        % etc.

\institute{Laboratory for Nuclear Science, Massachusetts Institute of Technology, 77 Mass. Ave., Cambridge, MA 02139, USA}

\abstract{Neutron stars exhibit magnetic fields and densities far beyond those achievable in terrestrial laboratories, offering a natural probe of strongly interacting matter under extreme conditions. Using observationally anchored mass-radius relations and a density profile consistent with established equations of state, we construct a piecewise model that explicitly integrates the neutron-drip line, nuclear-saturation, the electron-dominated halo, and core-crust interfaces. The resulting structure reproduces the stiffness and curvature behavior across the nuclear-pasta regime reported in the literature, validating our treatment of the crust-core transition. From this model, we derive updated moments of inertia, crustal mass fractions, and the effective number of neutrons contributing to the star's magnetic moment. Comparing these quantities with spin-down inferred magnetic dipole moments indicates that the observed magnetic fields of particularly millisecond pulsars can be sustained entirely by the crustal neutron polarization, requiring alignment of only about $\lesssim5.5\%$ ($99\%$ C.L.) of the neutrons in the crust. This finding supports a crust-confined magnetic-field origin for non-magnetar neutron stars, consistent with magneto-thermal evolution studies, and provides a quantitative framework for connecting neutron-star observables to its underlying structure.}

\maketitle

\tableofcontents

\section{Introduction}
\label{intro}

\begin{figure*}
\centering
\includegraphics[width=\textwidth]{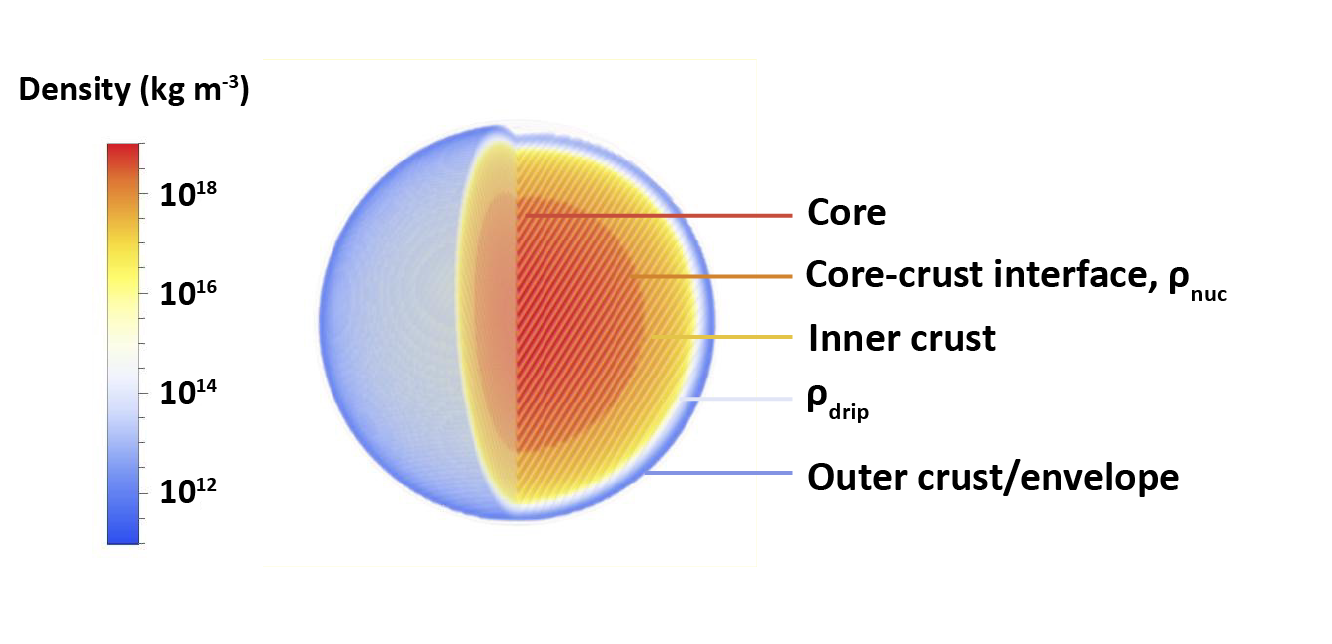}
\caption{Model of PSR J0437-4715 using the density model described in the text. This model uses a global charge neutrality condition, which can be seen with the mass discontinuity at the core-crust interface. Note the nuclear density $\rho_{nuc}$ marks the end of the core and the drip density $\rho_{drip}$ marks the end of the inner crust.}
\label{fig-1}
\end{figure*}

The neutron star equation of state is typically solved under a local charge-neutrality condition, in which the number density of protons and electrons is the same, \emph{i.e.} $n_p(r)=n_e(r)$ at every point $r$ within the neutron star, leading to a continuous density profile. Recent models have shown this violates the constancy of the Klein potentials and relaxed this condition to global neutrality, where the only requirement is that the neutron star as a whole is neutral, \emph{i.e.} $\int \rho_{ch}~ dV =0$ \cite{Belvedere2012-ng}. This allows an electron halo beyond the core-crust interface with a slightly proton-rich core. The electric field at the core-crust interface stores an energy density that consequently creates a jump in the mass-energy density. The discontinuity in the mass density alters mass-radius relations, as well as core radius estimates. Figure~\ref{fig-1} illustrates the density discontinuity at the core-crust interface within the model developed here.

The outer crust of neutron stars is the outermost layer of the star with densities less than the neutron drip density. The atoms here are stripped of electrons due to high pressure, and a crystalline lattice of neutron-rich nuclei is left. Free electrons are allowed to move through the nuclear matter. Approaching the inner crust, electron capture drives nuclei to be more neutron-rich \cite{Chamel2008-xg, Lorenz1993-gz}.

The inner crust of the neutron star lies between the neutron drip density, where free neutrons begin to leak out of nuclei, and the nuclear saturation density, where nuclei dissolve into uniform nuclear matter \cite{Negele1973-hd}. The inner crust contains neutron-rich nuclei, in which heavy nuclei are packed into a Coulomb lattice, and degenerate electrons provide charge neutrality and contribute to pressure outwards, balancing the gravitation pull of the star inwards \cite{Baym1971-wc, Haensel2006-uh}. Free neutrons here drip out of the nuclei and form a superfluid in the interstitial space of the lattice \cite{Watanabe2000-ku}. Near the core-crust interface, there is the nuclear-pasta state of matter where nuclei rearrange into exotic shapes due to the balance between nuclear attraction and Coulomb repulsion. Our knowledge gap in nucleon-nucleon interactions and solving quantum many-body calculations of nuclei creates uncertainties in the Equation of State (EoS) of the inner crust. %Changes in the structure of the inner crust have a large impact on transport properties and thus magnetic field evolution. 
The inner crust ultimately acts as a resistive, anisotropic layer that strongly shapes the magnetic field's long-term evolution \cite{Nandi2018-ut, Vigano2013-nh}.

The core of the neutron star consists of densities greater than the nuclear saturation density, and its contents, specifically of the inner core, are relatively unknown. While the microscopic theory of the core is not well constrained, the superconducting and superfluid properties of the dense matter within the core can be used to constrain the bulk magnetic behavior. Specifically, the presence of superconducting protons raises questions about the magnetic properties of neutron star cores \cite{Douchin2001-es}. Many neutron star models, specifically early models, claim that the superfluid neutrons and superconducting protons cannot support bulk polarization \cite{Vigano2013-nh}. Thus, the core does not significantly contribute to the magnetic dipole of a neutron star \cite{Cumming2004-bi}. 

However, observations of large magnetic field neutron stars, such as magnetars, suggest that the core may contribute to the field through fluxoids that are pinned to the proton superconductor. Further, purely crust-confined magnetic field configurations may not be capable of explaining the behavior of large magnetic field stars \cite{Wood2015-tg, Gourgouliatos2016-wx, Lander2019-ds}. However, the magnetic stresses and long-term stability of ordinary pulsars heavily differ from those of large magnetic field stars. The investigation of neutron star magnetic field origins and core contributions is ongoing \cite{Ho2017-gm, Gourgouliatos2022-kh, Pons2012-fb, Gusakov2020-rd}. In this paper, we rely primarily on stars with precise observational measurements, \emph{i.e.} millisecond pulsars, that do not possess extremely high magnetic fields. 

We return to earlier models that suggest that a crust-confined magnetic field was possible through Hall and Ohmic evolution. Our goal is not to exclude core magnetic fields generally, but rather to investigate if ordinary and millisecond pulsars are capable of originating the entirety of their magnetic field from the domains on their crusts \cite{Cumming2004-bi, Pons2012-fb}.

Neutron stars also currently provide some of the strongest constraints on axion and axion-like particle models, through their cooling \cite{Buschmann2022-ul}, spin, and surface magnetic field evolution \cite{Noordhuis2024-jq}. Our detailed treatment of the crustal structure and magnetic-field origins, therefore, not only constrains conventional neutron-star physics but also aids the astrophysical inputs entering these searches for physics beyond the Standard Model.

\section{Mass-radius relation}
The EoS models the relationship between pressure and energy density in a neutron star. Currently, there is not a complete EoS that describes all behaviors of the neutron star, but the Tolman-Oppenheimer-Volkoff (TOV) \cite{Tolman1939-xs, Oppenheimer1939-ab} equations are used to provide the relativistic generalization of hydrostatic equilibrium \cite{Lattimer2001-me, Chavez-Nambo2021-or}. The TOV equations are solved for different central densities and yield different neutron star models, with unique masses and radii. The mass-radius (M-R) relation is very sensitive to the EoS chosen.

Measurements of the mass and radius of neutron stars provide insight into the choice of EoS that matches observation; however, insights gained in this fashion are limited. Mass measurements are typically determined through pulsar timing in binary systems, which prevents the measurement of magnetar masses. Observing effects such as Shapiro delay \cite{Demorest2010-gn}, periastron advance \cite{Hulse1975-vd}, and orbital decay from gravitational wave emission allows rather precise mass measurements \cite{Weisberg2010-fj}. However, pulsar timing independently does not provide information on the radius of a star.

\begin{figure*}
\centering
\includegraphics[width=\textwidth]{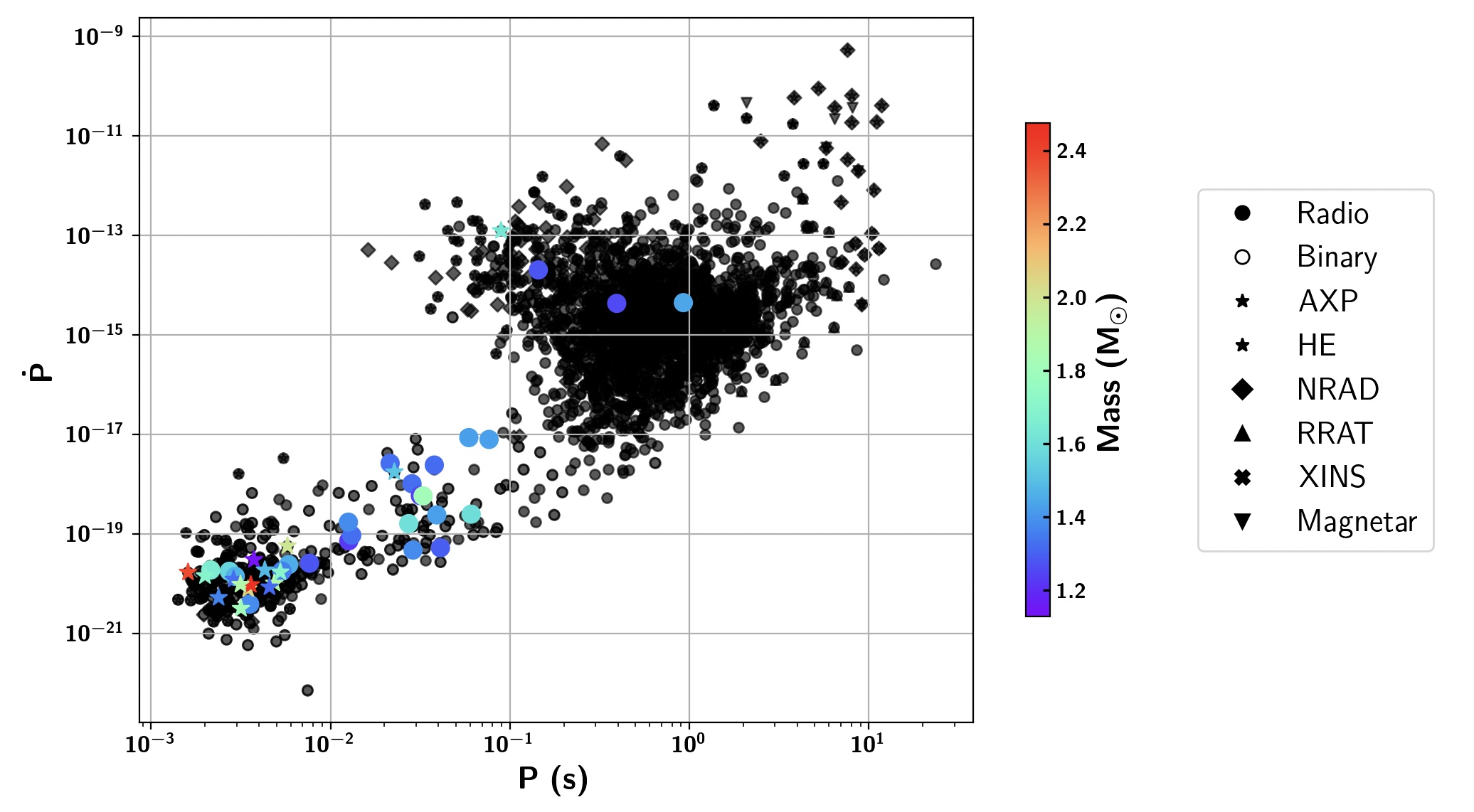}
\caption{Plot showing the period and the time-rate of change of the period for all stars in the ATNF catalog \cite{Manchester2005-vu} with their respective categories, noting that the majority of neutron stars with documented mass values, denoted by markers with color correlating to the color scale on the right, are considered to be millisecond pulsars.}
\label{fig-2} 
\end{figure*}

The radius is much more difficult to estimate and relies on X-ray observations \cite{Ozel2016-xe,Watts2016-ci,Lattimer2021-ys}. Thermal X-ray spectra of isolated neutron stars or X-ray binaries are fit with their atmosphere models to estimate the apparent radius \cite{Ozel2016-xe}. Newer techniques, such as pulse-profile modeling from the NICER mission, observe the relativistic light bending and how it affects the shape of pulsations in hot spots on the surface of neutron stars. The shape of these hot spots depends on the compactness (M/R), and sometimes this can be used to measure both the mass and radius \cite{Miller2019-ei, Miller2021-eg}. This limits investigations to ordinary pulsars, usually millisecond pulsars, which can be seen in the lower left corner of Figure \ref{fig-2}, as they have very stable, short periods that one can precisely measure. Conversely, large magnetic field stars are rarely found in binaries and present significant timing noise. Reliable constraints on the internal structure of neutron stars benefit from independent measurements of both their masses and radii, as this helps reduce dependence on assumptions of the EoS. A number of previous proposed EoS were subsequently ruled out when they failed to support stars with measured masses greater than $2M_{\odot}$ \cite{Demorest2010-gn,Lattimer2012-bh,Antoniadis2013-hi,Ozel2016-xe,Cromartie2019-bl}.

In this work, we therefore vet a meta-analyzed M-R relation constructed from such independent mass and radius determinations, and explicitly require consistency with representative EoS models discussed in the literature \cite{Ozel2016-xe,Watts2016-ci,Lattimer2021-ys,Lattimer2012-bh}. To anchor our density model, we then select four neutron stars with well-measured, independent mass and radius estimates, summarized in Table~\ref{tab-1}, and use these as reference points for the subsequent analysis.

\section{Neutron star density models}
\label{sec-1}
%yeah yeah yeah mention the methods of eos above and then here discuss why they are not great for this purpose

EoS models often have the largest uncertainties near the core-crust transition and densities above nuclear saturation, as the microscopic nature is not well understood. As a result, many density models in literature use local-neutrality single-branch with polytropes that smear the core-crust interface and miss (i) the interface energy jump from global neutrality \cite{Belvedere2012-ng}, (ii) stiffening across nuclear-pasta that modifies inner-crust curvature \cite{Negele1973-hd, Chamel2008-xg}, and (iii) neglect the independent observational values of $M$ and $R$ to constrain their respective density models \cite{Miller2019-ei, Miller2021-eg}. For testing if the large-scale dipole can be crust-confined in ordinary pulsars (or millisecond pulsars), a piecewise profile with explicit nuclear drip and saturation landmarks and a well-defined radius of the core, $R_{\rm core}$, is preferred.

\subsection{Density meta-model}

Millisecond pulsars (MSP) are the majority of stars with documented masses as seen in Figure~\ref{fig-2}. Thus, we use MSPs as the basis of our work. The central density and the average crust density were used as inputs into our model. For the central density, 8 different EoS models'(BPAL12 \cite{Ferrari2010-az}, BGN1H1 \cite{Potekhin2011-wq}, FPS \cite{Lorenz1993-gz}, BBB2 \cite{Datta1998-xa}, SLy \cite{Douchin2001-es}, BGN1 \cite{Potekhin2011-wq}, APR \cite{Akmal1998-gu}, BGN2 \cite{Potekhin2011-wq}) central density-mass relations were compared \cite{Haensel2006-uh}. These 8 EoS models are consistently used throughout literature and represent a wide range of stiffness and maximum mass values. The SLy (Skyrme Lyon) model is used as the midpoint since it contains a central density of approximately $1\times 10^{18}~\mathrm{kg~m^{-3}}$ and is near the average central density of the 8 models. Further, this model is often used for M-R relations and supports stars with a mass greater than two solar masses, $>2M_{\odot}$. The core here is modeled using the gravitational binding energy and the central value of the average density. Many other standard EoS do not support such high maximum masses \cite{Ozel2016-xe,Lattimer2016-vi,Haensel2004-xm}.

As mentioned in the introduction, the floor of the core is set at the nuclear saturation density of $\rho_{nuc} \approx 2.8\times 10^{17}~\mathrm{kg~m^{-3}}$, where the nuclear matter in the crust transitions away from nuclear-pasta structures. At this point, the sparse protons will form a superconductor, marking the beginning of the outer core \cite{Nandi2018-ut, Watanabe2000-ku, Yakovlev2015-dq}. Due to limited understanding of the nature of matter in the core of the neutron stars, our model does not reliably constrain structure deep within the star, below $2~$km. However, the mass contained within $2~$km only accounts for approximately 5$\%$ of the entire star's mass. Variations in the structure deep within the star therefore negligibly affects our estimations of the core radii, and as mentioned before, it similarly does not largely affect the star's magnetic field. 

\begin{figure*}[htbp]
\centering
\includegraphics[width=\textwidth]{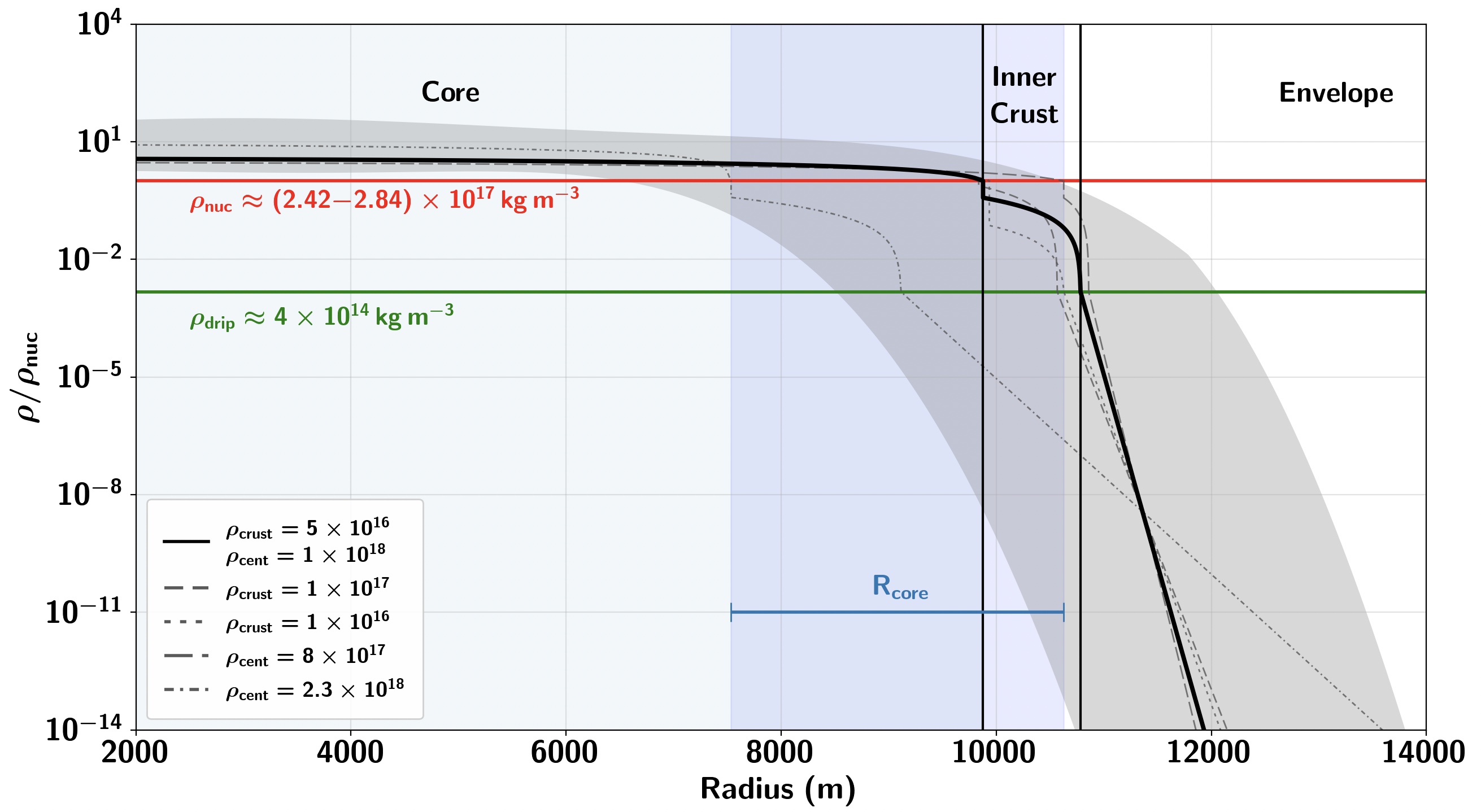}
\caption{Density model of J0437-4715 with documented mass $1.44\pm 0.07~M_{\odot}$ and radius $11.36\pm0.90$ km. The red line shows nuclear saturation, $\rho_{nuc}$ (ceiling core density) \cite{Douchin2001-es}, and the green line shows neutron drip, $\rho_{drip}$ (floor crust density) \cite{Negele1973-hd}. The dotted lines show the density model with the extremes of the central and average density values from the 8 different EoS models \cite{Haensel2006-uh}. The grey shaded region includes the propagated uncertainties from $\{M,R\}$ and density choices. The blue strip marks $\Rcore$ uncertainties within the $2\sigma$ range.}
\label{fig-3}
\includegraphics[width=\textwidth]{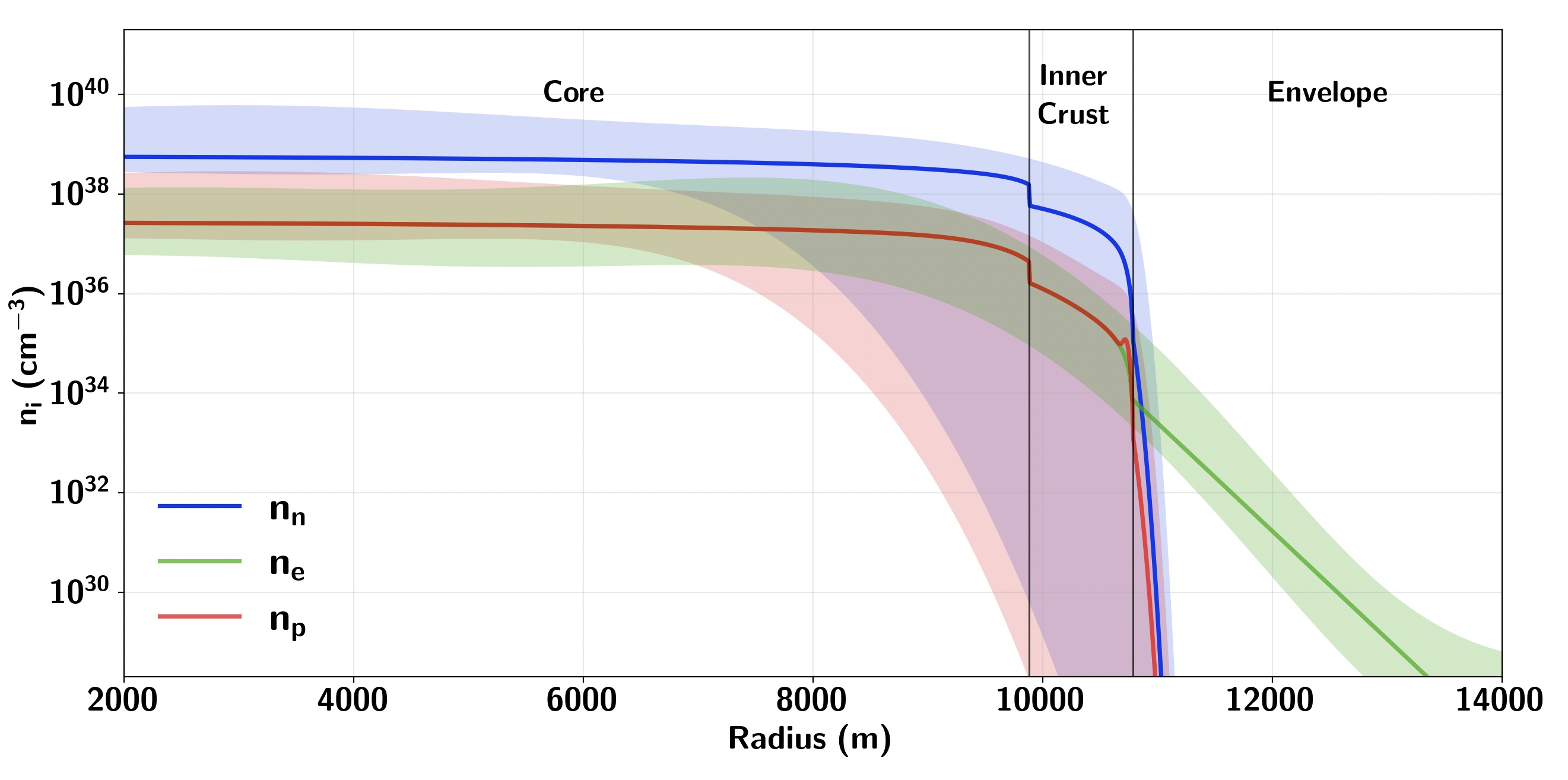}
\caption{Diagram depicting the neutron, proton, and electron density breakdown of the PSR J0437-4715 density model consistent with other global neutrality density models \cite{Belvedere2012-ng}.}
\label{fig-4}
\end{figure*}

The inner crust’s curvature is consistent with the fact that EoS continuously stiffens across nuclear-pasta phases\cite{Haensel2006-uh,Negele1973-hd}. The average crust density was used because there may be a density discontinuity at the core-crust interface, and this discontinuity may vary by star to preserve the global neutrality condition. To satisfy the local neutrality condition, the ceiling of core density can be as high as maximum nuclear density, $\rho_{\text{nuc}} = 2.8\times 10^{17}~\mathrm{kg~m^{-3}}$. In the case that the inner-core density ceiling was $\rho_{nuc}$, this would fulfill the local neutrality condition. The crust density floor is set at the neutron-drip density, $\rho_{drip}=4\times 10^{14}~\mathrm{kg~m^{-3}}$, because at this point the nuclei become neutron-rich such that free neutrons can be separated out of the nuclei. At densities lower than the drip density, this becomes the star's envelope, whose density decays exponentially. These densities, expressed piecewise can be written as \cite{Haensel2006-uh,Potekhin2011-wq,Douchin2001-es,Negele1973-hd}
\begin{equation}
\rho(r) = \begin{cases}
\Bigg(\rho_{0,~\rm core} \Bigg[1-\dfrac{G M r^2}{c^2R^4}\Bigg]^{a}\Bigg) r^4,~ r<R_{\rm core},\\
\left(b+c\ln{\rm r}\right) r^4,~R_{\rm core}<r<R_{\rm inner-crust},\\
r^4\rho_{\rm 0,~outer-crust} e^{-d\cdot r},~R_{\rm inner~crust}<r<R.
\end{cases}
\label{eq:rho}
\end{equation}
The individual terms in the above equation are explained and elaborated in the following paras. Eq.~\ref{eq:rho} is plotted in Figure \ref{fig-3}. As evident, we have assumed a spherically symmetric density.

\noindent\textit{Justification and precedent.}
Equation~\eqref{eq:rho} and Figure \ref{fig-3} show a piecewise, reduced-form model that (i) preserves the interface jump permitted by global neutrality \cite{Belvedere2012-ng}, (ii) encodes the well-known inner-crust curvature across nuclear-pasta \cite{Negele1973-hd, Watanabe2000-ku, Nandi2018-ut, Vigano2013-nh}, and (iii) treats the thin outer layers with an exponential scale characteristic of standard envelope/crust models \cite{Baym1971-wc, Gudmundsson1983-ky}. Using reduced-form models to match mass and radius ($\{M,R\}$) values along with interior trends is a standard practice (\emph{e.g.} using piecewise-polytrope fits) \cite{Lattimer2001-me,Read2009-vn}.

The core and crust branches in Eq.~\eqref{eq:rho} are joined at $R_{\rm core}$ with
(i) continuity of the enclosed mass $M(r=R_{\rm core})$, and (ii) allows for a small, upward density jump on the crust side motivated by global neutrality \cite{Belvedere2012-ng}. We enforce the integral mass constraint $\int_0^R 4\pi r^2 \rho(r)\,dr = M$ and anchor the landmarks at nuclear saturation and neutron-drip. The free shape parameters $(a,b,c,d)$ and normalizations (\emph{e.g.} $\rho_{0,\mathrm{core}}$) are determined by minimizing a residual that combines (i) the mass constraint, (ii) the core and inner crust density floor constraints, and (iii) the outer-envelope cutoff near $\rho \sim 10^9~\mathrm{kg\,m^{-3}}$ \cite{Haensel2006-uh}. At densities above $\sim 10^{9}~\mathrm{kg\,m^{-3}}$, the probability of electron capture on nuclei approaches one, rapidly driving the composition to more neutron-rich nuclei and marking the transition out of a light-element envelope \cite{Baym1971-wc, Gudmundsson1983-ky}. Dimensional consistency is ensured by absorbing scale factors into the fit coefficients; the explicit $r^4$ factors are a choice that anticipates the spherical shell Jacobian in later volume integrals (\emph{e.g.} the moment of inertia, $I$), and do not alter the physical content of the density profile.

Standard parametric models either use an analytic Tolman-type density profile for the core, joined smoothly to polytropic descriptions of the crust, or they approximate a full tabulated EoS using several polytropic segments fitted over different density ranges \cite{Read2009-vn, Lattimer2001-me}. Eq.~\eqref{eq:rho} differs in three ways that are helpful for the present goal: (1) it explicitly permits the small density jump at the core--crust boundary expected under global neutrality \cite{Belvedere2012-ng}; (2) it encodes the observed logarithmic
curvature across the inner crust with a weak $(b+c\ln r)$ dependence, rather than a single power law \cite{Negele1973-hd, Watanabe2000-ku, Nandi2018-ut, Vigano2013-nh}; and (3) it treats the thin outer layers with an exponential scale height in line with outer-crust/envelope physics \cite{Baym1971-wc, Gudmundsson1983-ky}. These choices keep the model compact, transparent, and directly anchored to $\{M,R\}$ while retaining the key microphysical landmarks.

\begin{table}
\small
\centering
\begin{tabular}{llll}
\hline
\hline
NS& Mass ($M_{\odot}$) & R (km) & $R_{core}$ (km)\\
\hline
J0437-4715\cite{Choudhury2024-ym} & $1.44\pm0.07$ & $11.4 \pm0.9$ & $9.65 \pm0.22$ \\
J0030+0451\cite{Miller2019-ei} & $2.01 \pm 0.15$ & $13.02 \pm 0.12$ & $11.03 \pm0.04$\\
J0348-0432 \cite{Huo2018-wd} & $1.44\pm0.04$ & $12.55 \pm 0.40$ & $9.56 \pm 0.05$\\
J0740+6620\cite{Miller2021-eg} & $2.08 \pm 0.07$ & $13.7 \pm 0.21$ & $11.04 \pm 0.11$\\
\hline
\hline
\end{tabular}
\caption{\label{tab-1}Four neutron stars (NS) that have both documented mass and radius values were used in our density model to find their core radius values. The same central and average densities were used for all 4 stars.}
\end{table}

\noindent\textit{Uncertainties and continuity.} The quoted mass and radius uncertainties in Table~\ref{tab-1} are taken directly from the original observational analyses for each source. The core-radius uncertainties are then obtained by propagating these errors associated with $\{M,R\}$ through our density model in Eq.~\ref{eq:rho}. The uncertainty associated with $\Rcore$ is set at the upper and lower global $1\sigma$ bounds of the mass-radius posteriors, as shown in Figure~\ref{fig-3}, and quoting the corresponding spread as the uncertainty.

Eq.~\ref{eq:rho}, written in a piecewise form, is not naturally continuous (or differentiable) at every interface. We allow for a small upward jump in the central value of the density at the core-crust boundary, motivated by models with global charge neutrality, which explicitly allow a discontinuity at the core-crust interface. This freedom is important because enforcing strict continuity there would artificially suppress a physically allowed feature of the EoS.

At the same time, the model is constructed so that, when the individual branches are combined into a full profile, its outer $1\sigma$ envelope smoothly interpolates between the inner and outer crust as well as the core. As illustrated in Figure~\ref{fig-3}, the resulting $1\sigma$ envelope of the density profile is adjusted to be effectively continuous by choosing appropriate values of the parameters $(a,b,c,d)$ and normalization constants in Eq.~\ref{eq:rho}. The central density value in the crust and its $1\sigma$ envelope (corresponding to $68.3\%$ confidence interval) is consistent with models in \emph{refs.} \cite{Douchin2001-es, Negele1973-hd}, but particularly with the models summarized in \emph{ref.}~\cite{Haensel2006-uh}. This smooth $1\sigma$ envelope ensures that volume-integrated quantities such as the enclosed mass, moment of inertia, and crustal mass fraction are well defined and numerically stable, while still retaining the possibility of a discontinuity at the core-crust boundary.

\begin{figure*}[h]
\centering
\includegraphics[width=\textwidth]{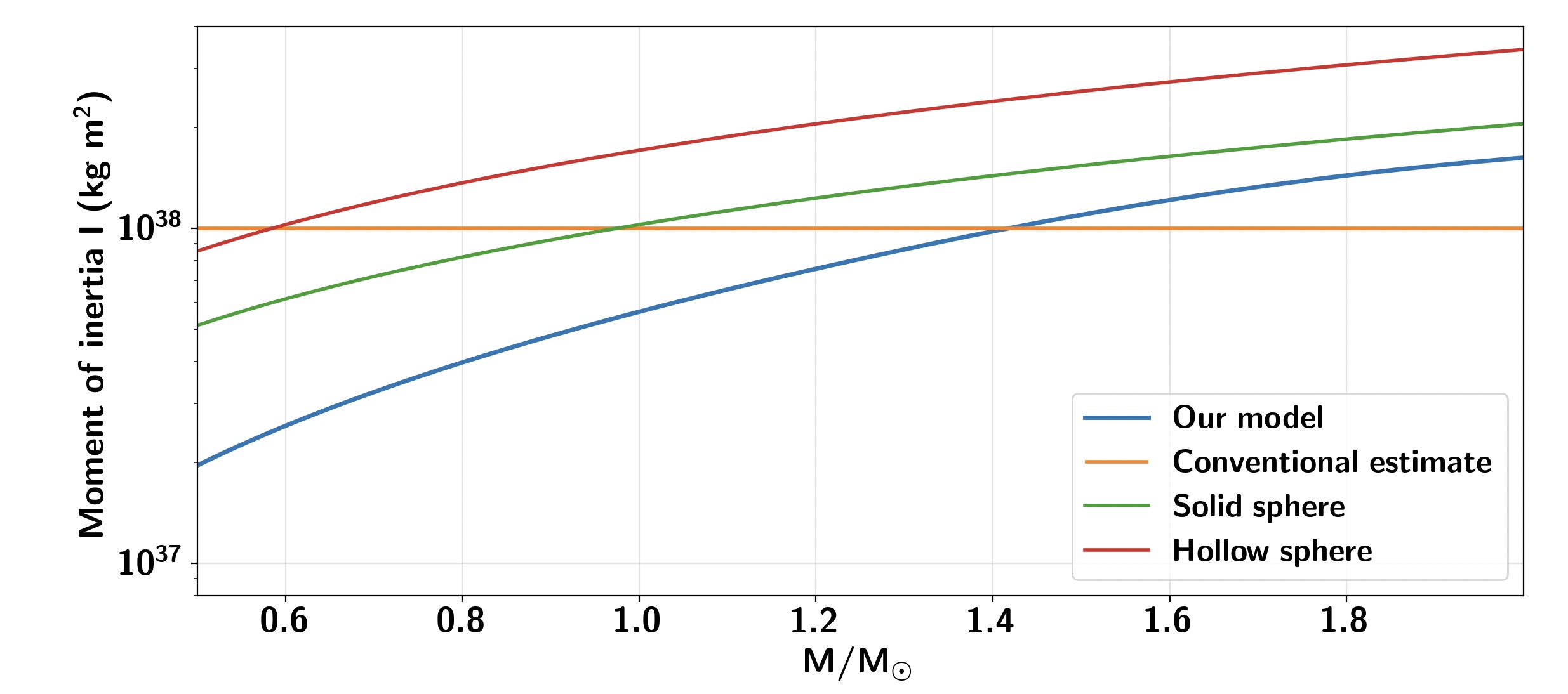}
\caption{Moment of inertia of a $11.36~$km star for masses varying from $ 0.5- 2~M_{\odot}$. This interval spans the full range of stable TOV solutions and includes all masses observed in real neutron stars, from the theoretical stability lower limit to the $\approx 2M_{\odot}$ maximal observed values. For reference, the canonical moment of inertia ($1\times 10^{38}~\rm kg~m^{-3}$ is shown in orange, as well as the moment of inertia for a solid and hollow sphere in green and red, respectively.}
\label{fig-5}
\end{figure*}

Once the total neutron density profile $\rho(r)$ is fixed by Eq.~\ref{eq:rho} and shown in Figure~\ref{fig-3}, the proton and electron densities were calculated similarly while respecting the global neutrality condition. The resulting density profiles for electrons and protons have also been plotted in Figure~\ref{fig-4}, and display similar key features described in the previous paras. However, the electron cloud extends far beyond the baryonic density profiles, supporting a positive core and inner crust, while ensuring global neutrality.

\subsection{Realistic moment of inertia}
With this density model, we can find the radius of the core and thus the thickness of the crust given a measured mass and radius, as seen in Figure~\ref{fig-2} and Table~\ref{tab-1}. The crust thickness is $\Delta R \equiv \Rstar - \Rcore$, where $R$ is the size of the entire neutron star and $\Rcore$ is the size of the neutron star's core. More specifically, the radius is defined by when the outer crust density becomes $10^9~\rm kg~m^{-3}$ \cite{Haensel2006-uh}. The crust thickness, coupled with the density model, allows one to know how many neutrons are within the crust. 

Further improvements were made by using the exact moment of inertia defined by
\begin{equation}
I=\int_0^R 4\pi r^4 \rho(r) dr,
\label{eq:I}
\end{equation}
where the density $\rho(r)$ is given by Eq.~\ref{eq:rho}. Canonically, a moment of inertia of $1\times 10^{38}~ \rm kg~m^2$ is used in \emph{refs.}~\cite{Haensel2006-uh, Lattimer2004-dt, Potekhin2005-vg}. However, as seen in Figure~\ref{fig-5}, this estimation is not accurate for stars with extreme masses. Given a more precise moment of inertia, the magnetic field values may also be improved. This aids our investigation into crustal domain polarization. Using the source-specific moment of inertia derived from Eq.~\eqref{eq:rho} rather than a canonical value incorporates the $I(M,R)$ variability expected across EoS \cite{Lattimer2001-me, Potekhin2005-vg}.

\section{Crust and magnetic field origins}
With the core radius values and thus the crust thickness, one can find the percentage of mass that originates from the crust as well as the number of neutrons in the crust, as shown in Figures~\ref{fig-3} and \ref{fig-4}. Further, given the magnetic dipole of the entire star and the total neutrons in the crust, one can find the percentage of neutrons in the star that are aligned with the star's magnetic dipole. In this section we have studied the magnetic field of neutron stars in light of independent measurements of their mass and radius, and in conjecture with the density model developed.

\noindent\textit{Formalism.} The rate of change of kinetic energy and surface magnetic field of a neutron star are given by \cite{Condon2016-qk}
\begin{eqnarray}
I\Omega\dot{\Omega} &=& -\frac{B_p^{2} R^6 \Omega^{4}}{6c^3}\sin^2\alpha,
~\text{where},~\nonumber\\
B_p &=& \sqrt{\frac{6c^3 I |\dot{\Omega}|}{R^6 \Omega^{3} \sin^2\alpha}},~
\label{eq:Bp}
\end{eqnarray}
where $\Omega$ is the angular velocity, $\Rstar$ is the size of the neutron star, $\alpha$ is the obliquity factor defined by the angle between the magnetic dipole moment and the spinning axis of the neutron star, and $I$ is the moment of inertia defined in Eq.~\ref{eq:I} and plotted in Figure~\ref{fig-5}. There are other magnetic field models, \emph{e.g.} the Force-free magnetospheres \cite{Potekhin2005-vg} and the axisymmetric magnetosphere \cite{Contopoulos1999-yd}.

The magnetic dipole moment, $\mu_{\rm NS}$, associated with a polar magnetic field $B_p$ and radius $R$ for a purely dipole configuration is given by \cite{Condon2016-qk,Camilo2000-iq}.
\begin{equation}
\mu_{\rm NS} = \frac{B_p R^3}{2}.\label{eq:mu}
\end{equation}
Knowing the total magnetic moment of the neutron star allows us to estimate the effective number of neutrons in crustal domains whose spins are mutually aligned.

The underlying microscopic origin of the neutron star's magnetic moment is the spins of the Fermions, dominated by neutrons, contained within it, as shown by their densities in Figure~\ref{fig-4}. The total number of neutrons in the crust is
\begin{equation}
N_{n,\mathrm{crust}} = \int_{R_{\rm core}}^{R} \frac{\rho(r)}{m_n}4\pi r^2 dr,~\label{eq:Nn}
\end{equation}
where, the density $\rho(r)$ is the neutron density in Eq.~\eqref{eq:rho}, and $m_n=1.674~927~498~04(95)\times10^{-27}~$kg \cite{Tiesinga2021-qm} is the mass of the neutron. Given the net total magnetic moment of the neutron star (Eq.~\ref{eq:mu}), and the number of neutrons in the crust (Eq.~\ref{eq:Nn}), in a purely dipolar model, the relative number of neutrons in the crustal domains whose spins are mutually aligned can be written as
\begin{equation}
\mathcal{P}_{\rm  spin} = \frac{\mu_{\rm NS}}{N_{n,\mathrm{crust}}|\mu_n|}.~\label{eq:Pol}
\end{equation}
It is important to note that $\mathcal{P}_{\rm spin}$ is effectively a book-keeping ratio we refer to as the \emph{spin-polarization}, whose value $\ll 1$ indicates a feasible crust-confined origins under the adopted assumptions.

\noindent\textit{Magnetosphere systematics.}
%Equation~\eqref{eq:Bp} uses the vacuum rotating-dipole normalization. 
Force-free magnetospheres modify the numerical prefactor and the $\alpha$-dependence at the tens-of-percent level \cite{Potekhin2005-vg}. Here, the rotating-dipole in a vacuum form is retained for a conservative, consistent comparison across sources; adopting a force-free calibration would shift $B_p$ without significantly changing the qualitative spin-polarization conclusions.

The stars' ages were found by comparing pulsars' right ascension, declination, and distance with the ages of nearby star clusters \cite{Kharchenko2005-tb}. The thresholds were set to be within 10 degrees for the right ascension and declination and 20 pc for the distance. Only pulsars with documented masses were used in this process, and the radius was found using the mass-radius relation from \emph{ref.} \cite{Belvedere2012-ng}. 

Globular clusters themselves form from a singular massive molecular cloud in a rather short time period \cite{Kruijssen2015-et, Portegies-Zwart2010-eg}. A neutron star within a cluster is a result of the evolution and inevitable core collapse supernova of a massive star within the cluster \cite{Kuranov2006-lg, Ivanova2008-li}. This does indicate that the age of the neutron stars within a cluster is slightly younger; however, the lifetime of the progenitor is short due to its extreme mass. There may be exceptions to this. In particular, star clusters may contain multiple star populations, or an unrelated star cluster may gravitationally capture a neutron star.

Although the age of stars in binary systems may usually be reliably computed, when present in star clusters, they may experience a nonuniform mass transfer that may affect these calculations, as was the case for PSR J0514 4002A \cite{Freire2007-le}. Further, many MSPs are considered to be recycled pulsars, in which their companion is a white-dwarf that is only born after the MSP accretes mass from a main-sequence companion.

\begin{figure*}[h]
\centering
\includegraphics[width=\textwidth]{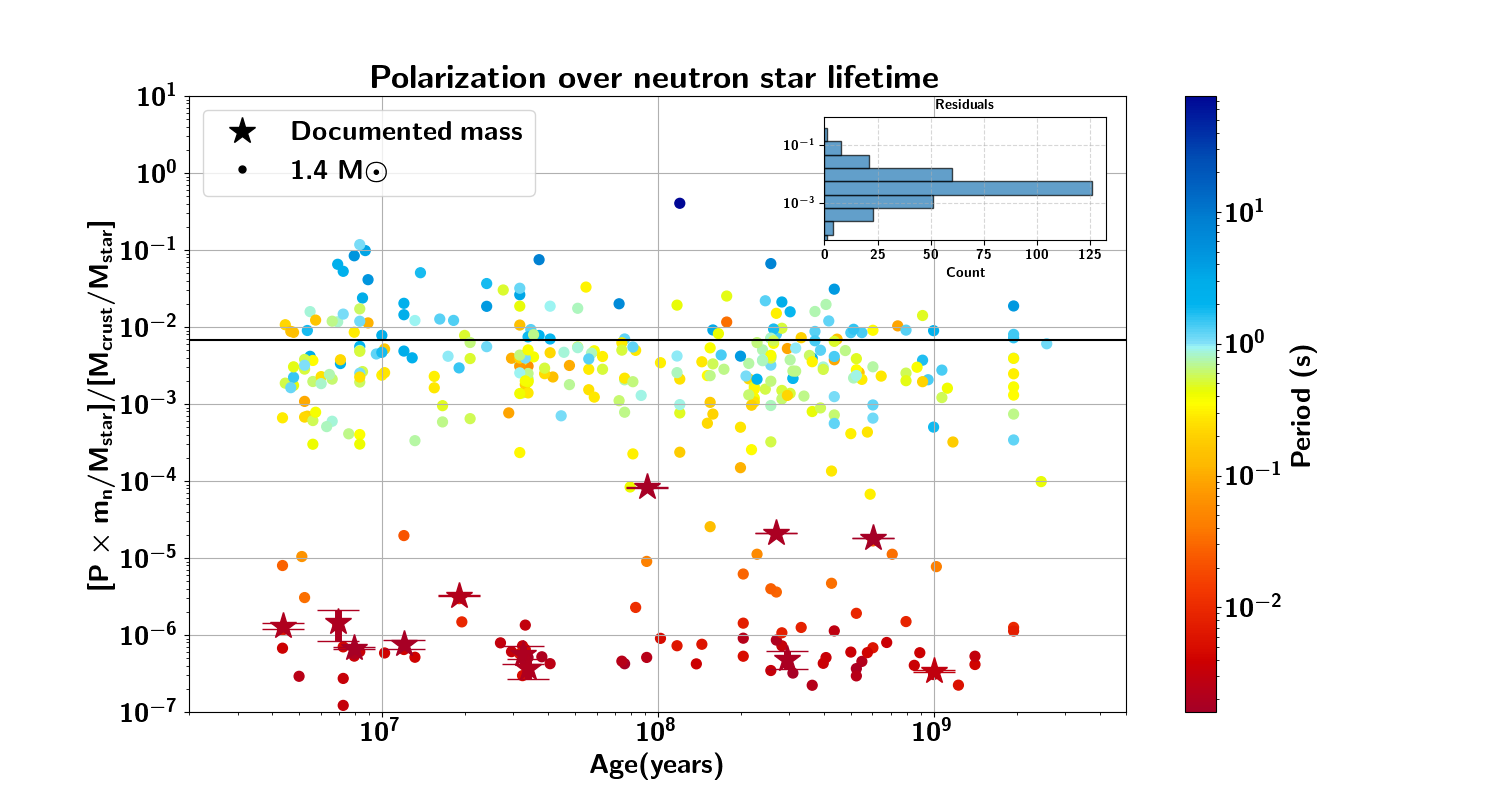}
\caption{Polarization for stars with documented mass, as well as all stars from the ATNF catalog \cite{Manchester2005-vu} with an assumed canonical mass. The age was derived by cross-checking the pulsars' right ascension, declination, and distance. The color scale on the right is to denote the relationship between period and polarization.}
\label{fig-6}
\end{figure*}

As seen in Figure~\ref{fig-6}, for stars with documented mass values (denoted by star markers), the ratio between the spin polarization mass and crustal mass remains significantly under one, showing it is feasible for these stars to originate their entire magnetic field within their crusts. However, we are limited to a small frame because, as mentioned earlier, most stars with documented mass are millisecond pulsars, and the polarization is proportional to the period of a star, following from Eqs.~\ref{eq:Bp}-\ref{eq:Pol}. %also age limitations

In order to further vet the spin-polarization of the crustal neutrons, for stars where we did not have independent mass and radius measurements, we used a canonical mass of $1.4 M_{\odot}$ \cite{Manchester2005-vu}, in combination with their period and rate of change of period. For all such stars, the ratio of the spin-polarization remained under one, suggesting that all of these stars are capable of generating the entirety of the star's magnetic field in their crust.

There is no apparent relationship between polarization and age, nor between period and age, but this may be due to limitations in age determination. The observable age range here, $\sim 10^7-10^9$ yr, restricts the range of magnetic field evolution we can observe. Further, work investigating core magnetic fields (see \emph{ref.} \cite{Ho2017-gm}), suggests that core magnetic fields remain effectively trapped in the star for $\sim 10^6-10^7$ years. Other works suggest that the magnitude and geometry of fields required to create a significant core contribution are unlikely in ordinary and millisecond pulsars \cite{Glampedakis2011, Passamonti2017}. In order to observe if these neutron stars are capable of containing their magnetic field for the entirety of their life, we must model the period over a neutron star's life. To do this in future investigations, one can look at the phase of the linearly polarized radio pulse \cite{Guillemot2023-jq}. This provides the magnetic moment independently and allows us to link the individualized spin-down rate for each star. 

As for the apparent gap between polarization values between $10^{-6}$ and $10^{-3}$, this can be explained by limitations in measurement rather than a consequence of spin-polarization behavior \cite{Miller2019-ei}. This is best visualized with the residuals diagram coupled with Figure~\ref{fig-6}, which demonstrates that the cluster around the ratio of $10^{-2}$ is consistent with a normal distribution. The gap in between this range of ratios corresponds to periods around $0.01~$s to $0.1~$s. Neutron stars in this period range are not often measured because frequencies associated with that period are very common man-made frequencies. During measurement, these frequencies are often filtered out as backgrounds, so as not to mistake these frequencies on Earth for a neutron star. The Parkes Survey, which heavily contributed to ATNF, demonstrated that the optimal sensitivity results from stars ranging from $0.1~\text{s}~\leq P \leq~2~$s \cite{Pons2012-fb, Manchester2001-qi}.

However, one can note a significant cluster in the millisecond range. This can be attributed to multiple measurement advantages presented by millisecond pulsars (MSPs) as well as their relatively long lifetime compared to other neutron stars \cite{Bhattacharya1991-sg}. MSPs are also very common in nearby globular clusters and more likely to be observed \cite{Camilo2000-iq}. Also, more recent surveys \cite{Cordes2006-fg, Abdo2009-mz, Keith2010-kf, Ray2011-bq}, unlike the Parkes survey, focused on optimizing MSP measurements. Further, MSPs produce bright gamma rays, and Fermi LAT has been a large contributor to matching MSPs to gamma ray sources \cite{Abdo2009-mz}. MSPs are approximately $20\%$ of documented neutron stars; however, they are predicted to be less than $5~\%$ of all neutron stars \cite{Lorimer2008-cq, Faucher-Giguere2006-ql}.

\noindent\textit{Uncertainties and limitations.}
The profile in Eq.~\eqref{eq:rho} is a surrogate anchored to $\{M,R\}$ and microphysical landmarks; it is not a full TOV solution for a single choice of EoS. As such, it is designed to capture the leading-order geometry relevant for a realistic moment of inertia and the crustal material. The error budget reported in Figure~\ref{fig-3} propagates uncertainties from $\{M,R\}$ and from the density-shape parameters in Eq.~\ref{eq:rho}. The global-neutrality jump is treated as a small allowed discontinuity at $R_{\rm core}$ rather than solved from a coupled Poisson-TOV system; this keeps the model agnostic to specific inner-core compositions while retaining the key features linked to physical landmarks described above. Finally, $\alpha$ is often not known for many stars. The spin-down inferences use the inclination $\alpha$-dependence explicitly, where $\alpha$ is unknown, a broad, reasonable prior is assumed, which widens the $B_p$ band but does not alter the finding that $\mathcal{P}_{\rm spin}\ll 1$ for the systems considered.

\section{Conclusion}
The majority of the stars provided are considered millisecond or ordinary pulsars; specifically, none of them are magnetars. This suggests that when not looking at magnetars, we may be able to model the magnetic field such that it is crust-exclusive. The fact established by Figure~\ref{fig-6}, where
\begin{equation}
\mathcal{P}_{\rm  spin} < 0.055~(99\%~\text{C.L.}),~\label{eq:c}
%0.007694780748264716+2.58*((0.03794770763676278-0.0014006815916083753)/2)
\end{equation}
which is indeed $\ll 1$ for the stars we have considered in this work, strongly supports the crustal origins of the magnetic moment of these neutron stars.

However, we recognize that there are many theories on the magnetic field origins and evolution of neutron stars, such as field-induced paramagnetism \cite{Peng2007-dv}, spontaneous ferro-crust or Landau-Stoner ferromagnetism \cite{Uma-Maheswari1997-au}, anisotropic $^3P_2$ superfluid magnetization \cite{Mizushima2021-hs}, chiral magnetic instability \cite{Dehman2025-xz}, spin-polarized ferromagnetic core phase \cite{Kutschera1999-mk}, and magnetized nuclear-pasta glass \cite{Yakovlev2015-dq}. The models that include a core magnetic field were formed with the behaviors of magnetars in mind. Our findings do not discredit other models, as their results reflect magnetar behavior better than the crust-exclusive models. However, when investigating neutron properties under extreme densities and magnetic fields, a crust-exclusive model allows for precise calculations, as the structure and contents of the crust are known much better than the core. This is significant, especially in nuclear astrophysical calculations, as millisecond and ordinary pulsars have independent mass and radius measurements, allowing for independent calculations on the properties of neutrons in the crust of these stars.

\section*{Acknowledgments}
We thank Prof. Robert Redwine for critical feedback regarding this work. A.L is supported by funding from NASA MA Space Grant Consortium, and P.M is supported by $\phi$K$\phi$ and ORAU fellowships.

%\clearpage

% References

\nocite{*}
%\bibliographystyle{aipnum-cp}%
%\renewcommand*{\bibfont}{\small}
%\bibliography{sample}%

\begin{thebibliography}{16cm}

\bibitem{Belvedere2012-ng}{R. Belvedere, D. Pugliese, J. A. Rueda, R. Ruffini, and S.-S. Xue, \emph{Neutron star equilibrium configurations within a fully relativistic theory with strong, weak, electromagnetic, and gravitational interactions}, Nucl. Phys. A {\bf 883}, 1 (2012). DOI: \href{https://doi.org/10.1016/j.nuclphysa.2012.02.018}{10.1016/j.nuclphysa.2012.02.018}.}

\bibitem{Chamel2008-xg}{N. Chamel and P. Haensel, \emph{Physics of neutron star crusts}, Living Rev. Relativ. {\bf 11}, 10 (2008). DOI: \href{https://doi.org/10.12942/lrr-2008-10}{10.12942/lrr-2008-10}.}

\bibitem{Lorenz1993-gz}{C. P. Lorenz, D. G. Ravenhall, and C. J. Pethick, \emph{Neutron star crusts}, Phys. Rev. Lett. {\bf 70}, 379 (1993). DOI: \href{https://doi.org/10.1103/PhysRevLett.70.379}{10.1103/PhysRevLett.70.379}.}

\bibitem{Negele1973-hd}{J. W. Negele and D. Vautherin, \emph{Neutron star matter at sub-nuclear densities}, Nucl. Phys. A {\bf 207}, 298 (1973). DOI: \href{https://doi.org/10.1016/0375-9474(73)90349-7}{10.1016/0375-9474(73)90349-7}.}

\bibitem{Baym1971-wc}{G. Baym, C. Pethick, and P. Sutherland, \emph{The ground state of matter at high densities: Equation of state and stellar models}, Astrophys. J. {\bf 170}, 299 (1971). DOI: \href{https://doi.org/10.1086/151216}{10.1086/151216}.}

\bibitem{Haensel2006-uh}{P. Haensel, A. Y. Potekhin, and D. G. Yakovlev, \emph{Neutron Stars: Equation of State and Structure}, 2007th ed., Springer - New York (2006). DOI: \href{https://doi.org/10.1007/978-0-387-47301-7}{10.1007/978-0-387-47301-7}.}

\bibitem{Watanabe2000-ku}{G. Watanabe, K. Iida, and K. Sato, \emph{Thermodynamic properties of nuclear ``pasta'' in neutron star crusts}, Nucl. Phys. A {\bf 676}, 455 (2000). DOI: \href{https://doi.org/10.1016/s0375-9474(00)00197-4}{10.1016/s0375-9474(00)00197-4}.}

\bibitem{Nandi2018-ut}{R. Nandi and S. Schramm, \emph{Transport properties of the nuclear pasta phase with quantum molecular dynamics}, Astrophys. J. {\bf 852}, 135 (2018). DOI: \href{https://doi.org/10.3847/1538-4357/aa9f12}{10.3847/1538-4357/aa9f12}.}

\bibitem{Vigano2013-nh}{D. Vigan\'o, N. Rea, J. A. Pons, R. Perna, D. N. Aguilera, and J. A. Miralles, \emph{Unifying the observational diversity of isolated neutron stars via magneto-thermal evolution models}, Mon. Not. R. Astron. Soc. {\bf 434}, 123 (2013). DOI: \href{https://doi.org/10.1093/mnras/stt1008}{10.1093/mnras/stt1008}.}

\bibitem{Douchin2001-es}{F. Douchin and P. Haensel, \emph{A unified equation of state of dense matter and neutron star structure}, Astron. Astrophys. {\bf 380}, 151 (2001). DOI: \href{https://doi.org/10.1051/0004-6361:20011402}{10.1051/0004-6361:20011402}.}

\bibitem{Cumming2004-bi}{A. Cumming, P. Arras, and E. Zweibel, \emph{Magnetic field evolution in neutron star crusts due to the Hall effect and ohmic decay}, Astrophys. J. {\bf 609}, 999 (2004). DOI: \href{https://doi.org/10.1086/421324}{10.1086/421324}.}

\bibitem{Wood2015-tg}{T. S. Wood and R. Hollerbach, \emph{Three dimensional simulation of the magnetic stress in a neutron star crust}, Phys. Rev. Lett. 114, 191101 (2015). DOI: \href{https://doi.org/10.1103/PhysRevLett.114.191101}{10.1103/PhysRevLett.114.191101}.}

\bibitem{Gourgouliatos2016-wx}{K. N. Gourgouliatos, T. S. Wood, and R. Hollerbach, \emph{Magnetic field evolution in magnetar crusts through three-dimensional simulations}, Proc. Natl. Acad. Sci. {\bf 113}, 3944 (2016). DOI: \href{https://doi.org/10.1073/pnas.1522363113}{10.1073/pnas.1522363113}.}

\bibitem{Lander2019-ds}{S. K. Lander and K. N. Gourgouliatos, \emph{Magnetic-field evolution in a plastically failing neutron-star crust}, Mon. Not. R. Astron. Soc. {\bf 486}, 4130 (2019). DOI: \href{https://doi.org/10.1093/mnras/stz1042}{10.1093/mnras/stz1042}.}

\bibitem{Ho2017-gm}{W. C. G. Ho, N. Andersson, and V. Graber, \emph{Dynamical onset of superconductivity and retention of magnetic fields in cooling neutron stars}, Phys. Rev. C 96, 065801 (2017). DOI: \href{https://doi.org/10.1103/PhysRevC.96.065801}{10.1103/PhysRevC.96.065801}.}

\bibitem{Gourgouliatos2022-kh}{K. Gourgouliatos, D. De Grandis, and A. Igoshev, \emph{Magnetic field evolution in neutron star crusts: Beyond the Hall effect}, Symmetry {\bf 14}, 130 (2022). DOI: \href{https://doi.org/10.3390/sym14010130}{10.3390/sym14010130}.}

\bibitem{Pons2012-fb}{J. A. Pons, D. Vigan\'o, and U. Geppert, \emph{Pulsar timing irregularities and the imprint of magnetic field evolution}, Astron. Astrophys. {\bf 547}, A9 (2012). DOI: \href{https://doi.org/10.1051/0004-6361/201220091}{10.1051/0004-6361/201220091}.}

\bibitem{Gusakov2020-rd}{M. E. Gusakov, E. M. Kantor, and D. D. Ofengeim, \emph{Magnetic field evolution time-scales in superconducting neutron stars}, Mon. Not. R. Astron. Soc. {\bf 499}, 4561 (2020). DOI: \href{https://doi.org/10.1093/mnras/staa3160}{10.1093/mnras/staa3160}.}

\bibitem{Buschmann2022-ul}{M. Buschmann, C. Dessert, J. W. Foster, A. J. Long, and B. R. Safdi, \emph{Upper limit on the QCD axion mass from isolated neutron star cooling}, Phys. Rev. Lett. {\bf 128}, 091102 (2022). DOI: \href{https://doi.org/10.1103/PhysRevLett.128.091102}{10.1103/PhysRevLett.128.091102}.}

%%%%%%%%%%%%%%%

\bibitem{Noordhuis2024-jq}{D. Noordhuis, A. Prabhu, C. Weniger, and S. J. Witte, \emph{Axion clouds around neutron stars}, Phys. Rev. X {\bf 14}, 041015 (2024). DOI: \href{https://doi.org/10.1103/physrevx.14.041015}{10.1103/physrevx.14.041015}.}

\bibitem{Tolman1939-xs}{R. C. Tolman, \emph{Static solutions of Einstein’s field equations for spheres of fluid}, Phys. Rev. {\bf 55}, 364 (1939). DOI: \href{https://doi.org/10.1103/physrev.55.364}{10.1103/physrev.55.364}.}

\bibitem{Oppenheimer1939-ab}{J. R. Oppenheimer and G. M. Volkoff, \emph{On Massive Neutron Cores}, Phys. Rev. {\bf 55}, 374 (1939). DOI: \href{https://doi.org/10.1103/PhysRev.55.374}{10.1103/PhysRev.55.374}.}

%%%%%%%%%%%%%%%%%

\bibitem{Lattimer2001-me}{J. M. Lattimer and M. Prakash, \emph{Neutron star structure and the equation of state}, Astrophys. J. {\bf 550}, 426 (2001). DOI: \href{https://doi.org/10.1086/319702}{10.1086/319702}.}

\bibitem{Chavez-Nambo2021-or}{E. Ch\'avez Nambo and O. Sarbach, \emph{Static spherical perfect fluid stars with finite radius in general relativity: a review}, Rev. Mex. Fis. E {\bf 18}, 020208 (2021). DOI: \href{https://doi.org/10.31349/revmexfise.18.020208}{10.31349/revmexfise.18.020208}.}

\bibitem{Demorest2010-gn}{P. B. Demorest, T. Pennucci, S. M. Ransom, M. S. E. Roberts, and J. W. T. Hessels, \emph{A two-solar-mass neutron star measured using Shapiro delay}, Nature {\bf 467}, 1081 (2010). DOI: \href{https://doi.org/10.1038/nature09466}{10.1038/nature09466}.}

\bibitem{Hulse1975-vd}{R. A. Hulse and J. H. Taylor, \emph{Discovery of a pulsar in a binary system}, Astrophys. J. {\bf 195}, L51 (1975). DOI: \href{https://doi.org/10.1086/181708}{10.1086/181708}.}

\bibitem{Weisberg2010-fj}{J. M. Weisberg, D. J. Nice, and J. H. Taylor, \emph{Timing measurements of the relativistic binary pulsar psr b1913+16}, Astrophys. J. {\bf 722}, 1030 (2010). DOI: \href{https://doi.org/10.1088/0004-637x/722/2/1030}{10.1088/0004-637x/722/2/1030}.}

\bibitem{Ozel2016-xe}{F. {\"O}zel and P. Freire, Masses, \emph{Radii, and the Equation of State of Neutron Stars}, Annu. Rev. Astron. Astrophys. {\bf 54}, 401 (2016). DOI: \href{https://doi.org/10.1146/annurev-astro-081915-023322}{10.1146/annurev-astro-081915-023322}.}

\bibitem{Watts2016-ci}{A. L. Watts et al., \emph{Colloquium: Measuring the neutron star equation of state using x-ray timing}, Rev. Mod. Phys. {\bf 88}, (2016). DOI: \href{https://doi.org/10.1103/revmodphys.88.021001}{10.1103/revmodphys.88.021001}.}

\bibitem{Lattimer2021-ys}{J. M. Lattimer, \emph{Neutron stars and the nuclear matter equation of state}, Annu. Rev. Nucl. Part. Sci. {\bf 71}, 433 (2021). DOI: \href{https://doi.org/10.1146/annurev-nucl-102419-124827}{10.1146/annurev-nucl-102419-124827}.}

\bibitem{Choudhury2024-ym}{D. Choudhury et al., \emph{A NICER view of the nearest and brightest millisecond pulsar: PSR J0437-4715}, Astrophys. J. Lett. {\bf 971}, L20 (2024). DOI: \href{https://doi.org/10.3847/2041-8213/ad5a6f}{10.3847/2041-8213/ad5a6f}.}

\bibitem{Miller2019-ei}{M. C. Miller et al., \emph{PSR J0030+0451 Mass and Radius from NICER Data and Implications for the Properties of Neutron Star Matter}, Astrophys. J. Lett. {\bf 887}, L24 (2019). DOI: \href{https://doi.org/10.3847/2041-8213/ab50c5}{10.3847/2041-8213/ab50c5}.}

\bibitem{Huo2018-wd}{J.-L. Huo and X.-F. Zhao, \emph{The moment of inertia of the proto-neutron star PSR J0348+0432 under neutrino trapped}, Chin. J. Phys. {\bf 56}, 292 (2018). DOI: \href{https://doi.org/10.1016/j.cjph.2017.11.027}{10.1016/j.cjph.2017.11.027}.}

\bibitem{Miller2021-eg}{M. C. Miller et al., \emph{The radius of PSR J0740+6620 from NICER and XMM-newton data}, Astrophys. J. Lett. {\bf 918}, L28 (2021). DOI: \href{https://doi.org/10.3847/2041-8213/ac089b}{10.3847/2041-8213/ac089b}.}

\bibitem{Lattimer2012-bh}{J. M. Lattimer, \emph{The Nuclear Equation of State and Neutron Star Masses}, Annu. Rev. Nucl. Part. Sci. {\bf 62}, 485 (2012). DOI: \href{https://doi.org/10.1146/annurev-nucl-102711-095018}{10.1146/annurev-nucl-102711-095018}.}

\bibitem{Antoniadis2013-hi}{J. Antoniadis et al., \emph{A massive pulsar in a compact relativistic binary}, Science {\bf 340}, 448 (2013). DOI: \href{https://doi.org/10.1126/science.1233232}{10.1126/science.1233232}.}

\bibitem{Cromartie2019-bl}{H. T. Cromartie et al., \emph{Relativistic Shapiro delay measurements of an extremely massive millisecond pulsar}, Nat. Astron. {\bf 4}, 72 (2019). DOI: \href{https://doi.org/10.1038/s41550-019-0880-2}{10.1038/s41550-019-0880-2}.}
  


\bibitem{Ferrari2010-az}{V. Ferrari, L. Gualtieri, and F. Pannarale, \emph{Neutron star tidal disruption in mixed binaries: The imprint of the equation of state}, Phys. Rev. {\bf 81}, (2010). DOI: \href{https://doi.org/10.1103/physrevd.81.064026}{10.1103/physrevd.81.064026}.}

\bibitem{Potekhin2011-wq}{A. Y. Potekhin, \emph{The physics of neutron stars}, Physics Uspekhi {\bf 53} , 1235 (2011). DOI: \href{https://doi.org/10.3367/UFNe.0180.201012c.1279}{10.3367/UFNe.0180.201012c.1279}.}

\bibitem{Datta1998-xa}{I. Bombaci, A. V. Thampan, and B. Datta, \emph{Rotating Neutron Stars for a New Microscopic Equation of State}, Stellar Astrophysics, Springer Netherlands (2000), pp. 405-410. DOI: \href{https://doi.org/10.1007/978-94-010-0878-5_48}{10.1007/978-94-010-0878-5\_48}.\\B. Datta, A. V. Thampan, and I. Bombaci, \emph{Equilibrium sequences of rotating neutron stars for new microscopic equations of state}, Astron. Astrophys. {\bf 334}, 943 (1998). \href{https://arxiv.org/abs/astro-ph/9801312}{arXiv: [astro-ph: 9801312]}.}

\bibitem{Akmal1998-gu}{A. Akmal, V. R. Pandharipande, and D. G. Ravenhall, \emph{Equation of state of nucleon matter and neutron star structure}, Phys. Rev. C {\bf 58}, 1804 (1998). DOI: \href{https://doi.org/10.1103/PhysRevC.58.1804}{10.1103/PhysRevC.58.1804}.}

\bibitem{Lattimer2016-vi}{J. M. Lattimer and M. Prakash, \emph{The equation of state of hot, dense matter and neutron stars}, Phys. Rep. {\bf 621}, 127 (2016). DOI: \href{https://doi.org/10.1016/j.physrep.2015.12.005}{10.1016/j.physrep.2015.12.005}.}

\bibitem{Haensel2004-xm}{P. Haensel and A. Y. Potekhin, \emph{Analytical representations of unified equations of state of neutron-star matter}, Astron. Astrophys. {\bf 428}, 191 (2004). DOI: \href{https://doi.org/10.1051/0004-6361:20041722}{10.1051/0004-6361:20041722}.}

\bibitem{Yakovlev2015-dq}{D. G. Yakovlev, \emph{Electron transport through nuclear pasta in magnetized neutron stars}, Mon. Not. R. Astron. Soc. {\bf 453}, 581 (2015). DOI: \href{https://doi.org/10.1093/mnras/stv1642}{10.1093/mnras/stv1642}.}

\bibitem{Gudmundsson1983-ky}{E. H. Gudmundsson, C. J. Pethick, and R. I. Epstein, \emph{Structure of neutron star envelopes}, Astrophys. J. {\bf 272}, 286 (1983). DOI: \href{https://doi.org/10.1086/161292}{10.1086/161292}.}

\bibitem{Read2009-vn}{J. S. Read, B. D. Lackey, B. J. Owen, and J. L. Friedman, \emph{Constraints on a phenomenologically parametrized neutron-star equation of state}, Phys. Rev. D {\bf 79}, 124032 (2009). DOI: \href{https://doi.org/10.1103/PhysRevD.79.124032}{10.1103/PhysRevD.79.124032}.}

\bibitem{Lattimer2004-dt}{J. M. Lattimer and M. Prakash, \emph{The physics of neutron stars}, Science {\bf 304}, 536 (2004). DOI: \href{https://doi.org/10.1126/science.1090720}{10.1126/science.1090720}.}

\bibitem{Potekhin2005-vg}{A. Y. Potekhin, V. Urpin, and G. Chabrier, \emph{The magnetic structure of neutron stars and their surface-to-core temperature relation}, Astron. Astrophys. {\bf 443}, 1025 (2005). DOI: \href{https://doi.org/10.1051/0004-6361:20053628}{10.1051/0004-6361:20053628}.}

\bibitem{Condon2016-qk}{J. J. Condon and S. M. Ransom, \emph{Essential Radio Astronomy}, Princeton University Press, Princeton, NJ (2016). URL: \href{https://www.cv.nrao.edu/~sransom/web/xxx.html}{cv.nrao.edu/~sransom/web/xxx.html}.}

\bibitem{Contopoulos1999-yd}{I. Contopoulos, D. Kazanas, and C. Fendt, \emph{The axisymmetric pulsar magnetosphere}, Astrophys. J. {\bf 511}, 351 (1999). DOI: \href{https://doi.org/10.1086/306652}{10.1086/306652}.}

\bibitem{Camilo2000-iq}{F. Camilo, D. R. Lorimer, P. Freire, A. G. Lyne, and R. N. Manchester, \emph{Observations of 20 millisecond pulsars in 47 tucanae at 20 centimeters}, Astrophys. J. {\bf 535}, 975 (2000). DOI: \href{https://doi.org/10.1086/308859}{10.1086/308859}.}

\bibitem{Tiesinga2021-qm}{E. Tiesinga, P. J. Mohr, D. B. Newell, and B. N. Taylor, \emph{CODATA recommended values of the fundamental physical constants: 2018}, Rev. Mod. Phys. {\bf 93}, 025010 (2021). DOI: \href{https://doi.org/10.1103/RevModPhys.93.025010}{10.1103/RevModPhys.93.025010}.}

\bibitem{Kharchenko2005-tb}{N. V. Kharchenko, A. E. Piskunov, S. R\"oser, E. Schilbach, and R.-D. Scholz, \emph{Astrophysical parameters of Galactic open clusters}, Astron. Astrophys. {\bf 438}, 1163 (2005). DOI: \href{https://doi.org/10.1051/0004-6361:20042523}{10.1051/0004-6361:20042523}.}

\bibitem{Kruijssen2015-et}{J. M. D. Kruijssen, \emph{Globular clusters as the relics of regular star formation in ‘normal’ high-redshift galaxies}, Mon. Not. R. Astron. Soc. {\bf 454}, 1658 (2015). DOI: \href{https://doi.org/10.1093/mnras/stv2026}{10.1093/mnras/stv2026}.}

\bibitem{Portegies-Zwart2010-eg}{S. F. Portegies Zwart, S. L. W. McMillan, and M. Gieles, \emph{Young massive star clusters}, Annu. Rev. Astron. Astrophys. {\bf 48}, 431 (2010). DOI: \href{https://doi.org/10.1146/annurev-astro-081309-130834}{10.1146/annurev-astro-081309-130834}.}

\bibitem{Kuranov2006-lg}{A. G. Kuranov, \emph{Neutron stars in globular clusters: Formation and observational manifestations}, Astro. Lett. {\bf 32}, 393 (2006). DOI: \href{https://doi.org/10.1134/S106377370606003X}{10.1134/S106377370606003X}.}

\bibitem{Ivanova2008-li}{N. Ivanova, C. O. Heinke, F. A. Rasio, C. Bassa, Z. Wang, A. Cumming, and V. M. Kaspi, \emph{Formation of Millisecond Pulsars in Globular Clusters}, AIP Conf. Proc. {\bf 983}, 442-447 (2008). DOI: \href{https://doi.org/10.1063/1.2900271}{10.1063/1.2900271}.}

\bibitem{Freire2007-le}{P. C. C. Freire, S. M. Ransom, and Y. Gupta, \emph{Timing the eccentric binary millisecond pulsar in NGC 1851}, Astrophys. J. {\bf 662}, 1177 (2007). DOI: \href{https://doi.org/10.1086/517904}{10.1086/517904}.}

\bibitem{Manchester2005-vu}{R. N. Manchester, G. B. Hobbs, A. Teoh, and M. Hobbs, \emph{The Australia telescope national facility pulsar catalogue}, Astron. J. {\bf 129}, 1993 (2005). DOI: \href{https://doi.org/10.1086/428488}{10.1086/428488}.}

\bibitem{Glampedakis2011}{K. Glampedakis, D. I. Jones, and L. Samuelsson,
\emph{Ambipolar diffusion in superfluid neutron stars}, Mon. Not. R. Astron. Soc. {\bf 413}, 2021 (2011). DOI: \href{https://doi.org/10.1111/j.1365-2966.2011.18278.x}{10.1111/j.1365-2966.2011.18278.x}.}

\bibitem{Passamonti2017}{A. Passamonti, T. Akg\"un, J. A. Pons, and J. A. Miralles, \emph{The relevance of ambipolar diffusion for neutron star evolution}, Mon. Not. R. Astron. Soc. {\bf 465}, 3416 (2017). DOI: \href{https://doi.org/10.1093/mnras/stw2936}{10.1093/mnras/stw2936}.}


\bibitem{Guillemot2023-jq}{L. Guillemot, I. Cognard, W. van Straten, G. Theureau, and E. G\'erard, \emph{Improving pulsar polarization and timing measurements with the Nan\c{c}ay Radio Telescope}, Astron. Astrophys. {\bf 678}, A79 (2023). DOI: \href{https://doi.org/10.1051/0004-6361/202347018}{10.1051/0004-6361/202347018}.}

\bibitem{Manchester2001-qi}{R. N. Manchester et al., \emph{The Parkes multi-beam pulsar survey - I. Observing and data analysis systems, discovery and timing of 100 pulsars}, Mon. Not. R. Astron. Soc. {\bf 328}, 17 (2001). DOI: \href{https://doi.org/10.1046/j.1365-8711.2001.04751.x}{10.1046/j.1365-8711.2001.04751.x}.}

\bibitem{Bhattacharya1991-sg}{D. Bhattacharya, \emph{Formation and evolution of binary and millisecond radio pulsars}, Phys. Rep. {\bf 203}, 1 (1991). DOI: \href{https://doi.org/10.1016/0370-1573(91)90064-s}{10.1016/0370-1573(91)90064-s}.}

\bibitem{Cordes2006-fg}{J. M. Cordes et al., \emph{Arecibo pulsar survey using ALFA. I. survey strategy and first discoveries}, Astrophys. J. {\bf 637}, 446 (2006). DOI: \href{https://doi.org/10.1086/498335}{10.1086/498335}.}

\bibitem{Abdo2009-mz}{A. A. Abdo et al., \emph{Detection of high-energy gamma-ray emission from the globular cluster 47 Tucanae with Fermi}, Science {\bf 325}, 845 (2009). DOI: \href{https://doi.org/10.1126/science.1177023}{10.1126/science.1177023}.}

\bibitem{Keith2010-kf}{M. J. Keith et al., \emph{The High Time Resolution Universe Pulsar Survey - I. System configuration and initial discoveries: HTRU - I. System configuration}, Mon. Not. R. Astron. Soc. {\bf 409}, 619 (2010). DOI: \href{https://doi.org/10.1111/j.1365-2910.1088/0067-0049/194/2/1766.2010.17325.x}{10.1111/j.1365-2966.2010.17325.x}.}

\bibitem{Ray2011-bq}{P. S. Ray et al., \emph{Precise $\gamma$-ray timing and radio observations of 17 Fermi $\gamma$-ray pulsars}, Astrophys. J. Suppl. Ser. {\bf 194}, 17 (2011). DOI: \href{https://doi.org10.1088/0067-0049/194/2/17}{10.1088/0067-0049/194/2/17}.}

\bibitem{Lorimer2008-cq}{D. R. Lorimer, \emph{Binary and millisecond pulsars}, Living Rev. Relativ. {\bf 11}, 8 (2008). DOI: \href{https://doi.org/10.12942/lrr-2008-8}{10.12942/lrr-2008-8}.}

\bibitem{Faucher-Giguere2006-ql}{C.-A. Faucher-Giguere and V. M. Kaspi, \emph{Birth and evolution of isolated radio pulsars}, Astrophys. J. {\bf 643}, 332 (2006). DOI: \href{https://doi.org/10.1086/501516}{10.1086/501516}.}

\bibitem{Peng2007-dv}{Q.-H. Peng and H. Tong, \emph{The physics of strong magnetic fields in neutron stars}, Mon. Not. R. Astron. Soc. {\bf 378}, 159 (2007). DOI: \href{https://doi.org/10.1111/j.1365-2966.2007.11772.x}{10.1111/j.1365-2966.2007.11772.x}.}

\bibitem{Uma-Maheswari1997-au}{V. S. Uma Maheswari, D. N. Basu, J. N. De, and S. K. Samaddar, \emph{Spin polarised nuclear matter and its application to neutron stars}, Nucl. Phys. A {\bf 615}, 516 (1997). DOI: \href{https://doi.org/10.1016/s0375-9474(97)00002-x}{10.1016/s0375-9474(97)00002-x}.}

\bibitem{Mizushima2021-hs}{T. Mizushima, S. Yasui, D. Inotani, and M. Nitta, \emph{Spin-polarized phases of P23 superfluids in neutron stars}, Phys. Rev. C {\bf 104}, (2021). DOI: \href{https://doi.org/10.1103/physrevc.104.045803}{10.1103/physrevc.104.045803}.}

\bibitem{Dehman2025-xz}{C. Dehman and J. A. Pons, \emph{Magnetar field dynamics shaped by chiral anomalies and helicity}, Phys. Rev. Res. {\bf 7}, 033231 (2025). DOI: \href{https://doi.org/10.1103/rhv5-nd4v}{10.1103/rhv5-nd4v}.}

\bibitem{Kutschera1999-mk}{M. Kutschera, \emph{Emergence of magnetic field due to spin polarized baryon matter in neutron stars}, Mon. Not. R. Astron. Soc. {\bf 307}, 784 (1999). DOI: \href{https://doi.org/10.1046/j.1365-8711.1999.02655.x}{10.1046/j.1365-8711.1999.02655.x}.}

\end{thebibliography}

\end{document}